\documentclass[letter, twocolumn]{aa}
\usepackage{epsfig}
\usepackage{amsmath}   
\usepackage{amssymb}

\begin{document}

\title{A Qualitative Interpretation of the Second Solar Spectrum of Ce~ll}

\titlerunning{Second Solar Spectrum of Ce~{\sc ii}}

\author{R. Manso Sainz\inst{1} \and E. Landi Degl'Innocenti\inst{2}
\and J. Trujillo Bueno\inst{3, 4}}

\institute{High Altitude Observatory, National Center for
Atmospheric Research\thanks{The National Center for Atmospheric
Research is sponsored by the National Science Foundation.}, P.O. Box 3000, 
Boulder CO 80307-3000, USA
\and
Dipartimento di Astronomia e Scienza dello Spazio, 
Universit\`a degli Studi di Firenze, Largo Enrico Fermi 2,
I-50125, Firenze, Italy.
\and
Instituto de Astrof\'{\i}sica de Canarias, Av. V\'{\i}a L\'actea s/n,
E-38205, La Laguna, Tenerife, Spain.
\and
Consejo Superior de Investigaciones Cient\'\i ficas, Spain.}
\date{Received  / Accepted }

\abstract{This is a theoretical investigation on the formation of the 
linearly polarized line spectrum of ionized cerium in the sun.
We calculate the scattering line polarization pattern emergent from a 
plane-parallel layer of Ce~{\sc ii} atoms 
illuminated from below by the photospheric radiation field, 
taking into account the differential pumping 
induced in the various
magnetic sublevels by the anisotropic radiation field.
We find that the line polarization pattern calculated with this simple model 
is in good qualitative agreement with reported observations.
{Interestingly, the agreement improves when some amount 
of atomic level depolarization is considered. We find that
the best fit to the observations corresponds to 
the situation where the ground and metastable levels are
depolarized to about one fifth of the corresponding value obtained 
in the absence of any depolarizing mechanism. 
One possibility to have this situation is that the depolarizing rate
value of elastic collisions is exactly 
$D=10^6{\rm s}^{-1}$, which is rather unlikely. Therefore, we
interpret that fact as} due to the presence of a turbulent 
magnetic field in the limit of saturated Hanle effect for the lower-levels.
For this turbulent magnetic field we obtain a lower limit 
of 0.8 gauss and an upper limit of 200--300 gauss.
}

\maketitle

\section{Introduction}

When observed towards the limb, the solar spectrum is linearly polarized 
and it shows a surprisingly complex polarization structure, 
quite different from what might be intuitively expected from the
standard Fraunhofer spectrum 
(Stenflo \& Keller 1996, 1997).

One of the striking characteristics of this so-called second solar spectrum
is the preeminence of polarization signals from several
minority species.
For example, Ti~{\sc i} and Cr~{\sc i} have several multiplets and 
isolated lines that show relatively high fractional polarization values;
the same is true for several, innocent looking, lines of rare earth elements
that are hardly noticeable in the Fraunhofer spectrum.
This paper deals with the formation of the very interesting second solar
spectrum of a rare earth: Ce {\sc ii}.

Rare earths have configurations with more than one open shell
that yield a rich and complex level structure.
Ce {\sc ii} is an extreme example;
its two lower configurations $4f5d6s$ and $4f5d^2$ 
produce 46 levels within 6000 cm$^{-1}$ from the ground level 
(Martin, Zalubas, \& Hagan 1978), and 96 levels within 10000 cm$^{-1}$.
Such a wealth of low-lying levels 
gives rise to a large number of resonant or 
quasi-resonant transitions, and the population of cerium spreads 
among many metastable levels.
This fact, and the relatively low abundance of the element
($\alpha_{\rm Ce}=1.58$ in the usual logarithmic scale with 
$\alpha_{\rm H}=12$; Asplund, Grevesse \& Sauval 2005),
makes resonant lines of cerium form deep in the atmosphere.

An estimate of the height of formation $H$ 
at heliocentric angle $\theta$
of the center of a typical resonance line of  Ce {\sc ii} 
(the main ionization stage 
for solar atmospheric conditions),
is obtained from the expression
\begin{equation}
\frac{c^2}{8\pi^{3/2}\nu^2} \frac{A_{u\ell}}{\Delta\nu_D} 
\frac{g_u}{u(T)} 10^{(\alpha-12)} 
{\cal N}_0 \int^\infty_H \exp{(-z/{\cal H})} \frac{{\rm d}z}{\cos\theta}
\approx 1,
\end{equation}
where we use standard notation and ${\cal N}_0$ is the number density of hydrogen atoms at the reference
height $z=0$.
For ionized cerium, the degeneracy of a typical excited level 
radiatively connected to the ground level is
$g_u=8$ and the partition function at the effective temperature of the 
sun is $u(T=5800 {\rm K})\approx 230$
(van Diest 1980; Irwin 1981).
Therefore, a typical visible line 
(say, at wavelegth $\lambda\approx 5000$~\AA\ ,
with Doppler width $\Delta\nu_D 
\approx 1.65\times 10^9$~s$^{-1}$
and Einstein coefficient for spontaneous decay
$A_{ul}\approx 10^8$~s$^{-1}$), in an exponentially stratified atmosphere
with a scale height ${\cal H}\approx 130$ km and 
${\cal N}_0 = 1.2\times 10^{17}$ cm$^{-3}$ (Vernazza,
Avrett, \& Loeser 1976), forms at 
$H\approx {\cal H}\log(7/\mu)$ km. This is 
$H\approx 250$ km at disc center, and $H\approx$550 km near the limb at $\mu=0.1$, i.e.,
270 km below and 30 km above the photospheric minimum of temperature {of the classical one-dimensional semi-empirical models}, respectively.

\section{Scattering Line Polarization in Ce {\sc ii}}

An elementary analysis of scattering line polarization
in Ce {\sc ii} may be done along the lines of Manso Sainz \&
Landi Degl'Innocenti (2002, 2003; Papers {\sc i}a,~b from now on) 
for Ti {\sc i}.
We consider the excitation state of the atoms in a layer 
located at the top of the solar atmosphere and 
illuminated from below by the photospheric radiation field.
Furthermore, we assume that the radiation field is axisymmetric 
with respect to the local vertical
and that the polarization degree is very low (quiet sun case).
The excitation state of the
atoms is then described 
by the independent populations $N_M$ of the $(2J+1)$ magnetic
sublevels of each atomic level having total angular momentum 
$J$, where the quantization axis for angular momentum is chosen along the symmetry axis\footnote{We point out that cerium has no 
nuclear spin and hence, no hyperfine structure.}.
An equivalent and more convenient description may be given in terms of the elements (statistical tensors)
\begin{equation}
  \rho^K_0\,=\,\sum_{M=-J}^{J}(-1)^{J-M}\sqrt{2K+1}\left(
	\begin{array}{ccc}
	J & J & K \\
	M & -M & 0 
	\end{array}\right)\,{N_M},
								\label{3}
\end{equation}
where $K=0, 2, ..., 2J-1$ (Ce {\sc ii} levels have half-integer total
angular momentum). Note that $\rho^0_0$ 
is proportional to the total population of the level, while
$\rho^2_0$ is the so-called alignment of the level;
note also that odd-$K$ elements vanish since no population imbalance 
between $+M_J$ and $-M_J$ sublevels can be created under the hypothesis
in this research.
As will become apparent shortly, the emission, absorption and 
dichroism coefficients acquire a very simple and transparent form 
once expressed in terms of the $\rho^K_Q$ elements.

We consider a fairly complete atomic model of
Ce~{\sc ii} which has been compiled from the 
Database on Rare Earths at Mons University 
(DREAM\footnote{{\tt http://www.umh.ac.be/\lower3pt\hbox{\char'176}astro/dream.shtml}}
; Palmeri et al. 2000; Zhang et al. 2001).
This model has 500 levels and accounts for 16033 radiative transitions 
between them.
The levels have total angular momenta spanning from $J=1/2$ to $15/2$, and in
all 2038 $\rho^K_0$ elements are necessary to describe
the excitation state of the atom. 

Given an arbitrary level $i$ with total angular momentum $J$, 
the relaxation rate of the statistical tensor $\rho^K_0(i)$
due to spontaneous emission from level $i$ towards lower levels $\ell$ is 
(see Landi Degl'Innocenti \& Landolfi 2004)
\begin{equation}
-\rho^K_0(i) \sum_\ell A_{i\ell},            \label{see1}
\end{equation}
while the transfer rate due to spontaneous emission into the same level from 
upper levels $u$ with total angular momentum $J_u$ is
\begin{equation}
+\sum_{u}\rho^K_0(u) (2J_u+1)A_{ui}
	(-1)^{1+J+J_u}
	\left\{\begin{array}{ccc}J_u & J_u & K\\J & J&1\end{array}\right\}.
	                                       \label{see2}
\end{equation}
Similarly, the relaxation rate of $\rho^K_0(i)$ due to absorptions
from level $i$ towards upper levels $u$ is
\begin{equation}
\begin{split}
-\sum_{K'}\rho^{K'}_0(i)&(2J+1)\sum_{uK_u}B_{iu}\\
        &\times\sum_{K_{\rm r}=0, 2}
	\sqrt{3(2K+1)(2K'+1)(2K_{\rm r}+1)} \\
	&\times (-1)^{1+J_u-J}
	\left\{\begin{array}{ccc}
	K & K' & K_{\rm r} \\
	J & J & J \end{array}\right\}
	\left\{\begin{array}{ccc}
	1 & 1 & K_{\rm r} \\
	J & J & J_u \end{array}\right\} \\
	&\times\left(\begin{array}{ccc}
	K & K' & K_{\rm r} \\
	0 & 0 & 0 \end{array}\right) \! J^{K_{\rm r}}_0(u\rightarrow i),
	                                       \label{see3}
\end{split}
\end{equation}
and the transfer rate due to absorptions from
lower levels $\ell$ to level $i$, is
\begin{equation}
\begin{split}
+\sum_{\ell K_\ell} \rho^{K_\ell}_0(\ell)&(2J_\ell+1)B_{\ell i} \\
&\times\sum_{K_{\rm r}=0, 2}\sqrt{3(2K+1)(2K_\ell+1)(2K_{\rm r}+1)} \\
	&\times
	\left\{\begin{array}{ccc}
	J & J_\ell & 1 \\
	J & J_\ell & 1 \\
	K & K_\ell & K_{\rm r} \end{array}\right\}
	\left(\begin{array}{ccc}
	K & K_\ell & K_{\rm r} \\
	0 & 0 & 0 \end{array}\right)\!J^{K_{\rm r}}_0(i\rightarrow \ell).
	                                       \label{see4}
\end{split}
\end{equation}
In Eqs.~(\ref{see3})-(\ref{see4}), $J^0_0$ and $J^2_0$ 
are spherical components of the tensor that
describes the 
radiation field illuminating the atoms.
Neglecting the role of the (typically low) degree of polarization 
of the radiation field on
the population of atomic sublevels they read:
\begin{align}
  J^0_0&=\,\frac{1}{2}\int_{-1}^1 {\rm d}\mu' I(\mu'),     \label{1}
\\
  J^2_0&=\,\frac{1}{4\sqrt{2}}\int_{-1}^1 
	{\rm d}\mu' [(3\mu'{}^2-1)I(\mu')],        \label{2}
\end{align}
where $I(\mu')$ is the intensity and $\mu'$ the cosine of the angle that
the radiation beam forms with the vertical.

The balance of Eqs.~(\ref{see1})-(\ref{see4}) 
establishes the excitation state of the atom
in statistical equilibrium under a given illumination (stimulated emission
can be safely neglected and 
has not been considered). 
The result is an algebraic system of equations in the unknowns $\rho^K_0$.
This system is transformed into a  non singular one by substituting one of the redundant 
equations (for example, 
the equation for $\rho^0_0$ of the ground level), 
with the equation expressing the 
conservation of particles $\sum_i\rho^K_0(i)\sqrt{2J_i+1}=1$.

The intensity and linear polarization emissivities, 
along a $\mu$ direction, of a spectral line 
corresponding to the decay from a level $u$ to a level $\ell$ 
may be simply expressed as
\begin{align}
\begin{split}
  \epsilon^{\rm line}_I&=\,\epsilon_0 \phi_\nu [\rho^0_0(u)+
	w_{J_uJ_\ell}^{(2)} \frac{1}{2\sqrt{2}} (3\mu^2-1)\rho^2_0(u)],
\end{split}
								\label{9}
\\
\begin{split}
  \epsilon^{\rm line}_Q&=\,\epsilon_0 \phi_\nu w_{J_uJ_\ell}^{(2)} \frac{3}{2\sqrt{2}}(1-\mu^2)
	\rho^2_0(u),  \label{10}
\end{split}
\end{align}
where $\epsilon_0=(h\nu/4\pi)~A_{u\ell}\sqrt{2J_u+1}\,\,N$ 
($N$ being the number density of Ce~{\sc ii} atoms), $\phi_\nu$ is the emission profile and
$w_{J_uJ_\ell}^{(2)}$ is a numerical factor that depends only on the 
quantum numbers of the levels involved in the transition (Table 1
in Landi Degl'Innocenti \cite{gidi84} gives explicit values).
Reciprocally, in an optically thick medium absorption
($\eta_I$) and dichroism ($\eta_Q$) coefficients must also be considered.
They depend on the population and alignment of the lower level $\ell$
\begin{align}
\begin{split}
  \eta^{\rm line}_I&=\,\eta_0 \phi_\nu [\rho^0_0(\ell)+
	w_{J_\ell J_u}^{(2)} \frac{1}{2\sqrt{2}} (3\mu^2-1)\rho^2_0(\ell)],
\end{split}
								\label{11}
\\
\begin{split}
  \eta^{\rm line}_Q&=\,\eta_0 \phi_\nu w_{J_\ell J_u}^{(2)} \frac{3}{2\sqrt{2}}(1-\mu^2)
	\rho^2_0(\ell),       \label{11b}
\end{split}
\end{align}
where 
$\eta_0=(h\nu/4\pi)~B_{\ell u}\sqrt{2J_\ell+1}\;N$.

As in Papers {\sc i} we use a simplified radiative transfer model for the
formation of the scattering polarization signal.
The radiation illuminating the Ce~{\sc ii} ions is approximated by a
continuum (spectral details are neglected),
and expressions (\ref{1})-(\ref{2}) are evaluated from the observed 
solar center-to-limb variation of the continuum intensity (Cox 2000). 
This approximation is entirely justified 
because of the weakness of cerium lines in the intensity solar spectrum.
The emergent fractional polarization is then calculated in the 
limit of tangential
observation, which gives
\begin{equation}
   \left(\frac{Q}{I}\right)_{\mu\rightarrow 0^+}\,=\,
	\frac{\epsilon_Q/\epsilon_I-\eta_Q/\eta_I}
	{1-(\eta_Q/\eta_I)(\epsilon_Q/\epsilon_I)},    \label{12}
\end{equation}
where $\epsilon_i$ and $\eta_i$, which must be evaluated at the surface of the
plane-parallel slab, have contributions from both, continuum and 
spectral line.
Expression (\ref{12}) may be further simplified taking into account 
the low polarization level ($\eta_Q/\eta_I\ll 1$,
$\epsilon_Q/\epsilon_I\ll 1$). Thus, to first order in
$\epsilon_Q/\epsilon_I$ and $\eta_Q/\eta_I$, the fractional polarization
reduces to {(Trujillo Bueno 2003)}
\begin{equation}
   \frac{Q}{I}\,\approx\,
	\frac{\epsilon_Q}{\epsilon_I}-\frac{\eta_Q}{\eta_I}. \label{13}
\end{equation}

\begin{figure}
\centerline{\epsfig{figure=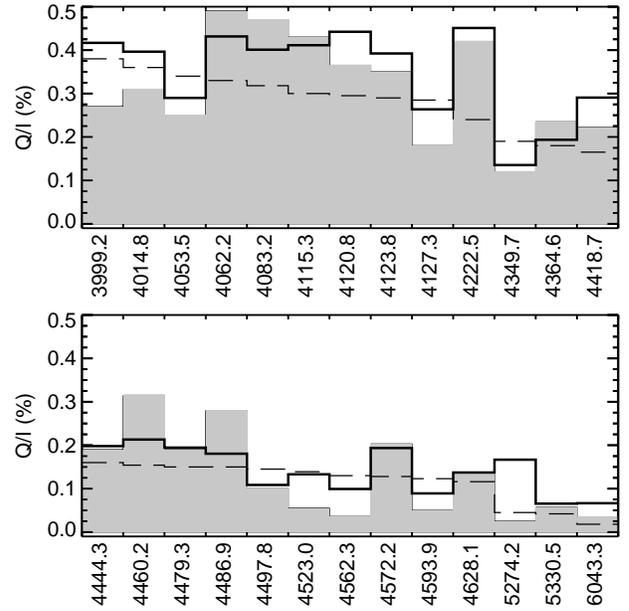, width=9cm}}
\caption{Main polarization features in the visible line spectrum
	of Ce~{\sc ii}. Shadowed bars and dashed line represent,
	respectively, the fractional
	polarization in the spectral lines 
	and in the adjacent continuum
	observed near the solar limb ($\mu=0.1$), as reported 
	by Gandorfer (2000, 2002). The heavy line represents the 
	fractional polarization as calculated in this work
	($\mu\rightarrow0$).
	For representation purpouses the calculated fractional
	polarization has been scaled by an arbitrary factor $\alpha=0.08$ and
	added to the continuum fractional polarization 
	(see Eq.~(\ref{14}) with $\eta_I^{\rm line}/\eta_I^{\rm cont}=\alpha$).}
\end{figure}

Expression (\ref{13}) may be further developed taking into account  
that rare earths show very weak lines in the solar spectrum.
Thus, in the limit $\eta_I^{\rm line}/\eta_I^{\rm cont}\ll 1$, 
Eq.~(\ref{13}) leads to
\begin{equation}
\frac{Q}{I}\,\approx\, \left(\frac{Q}{I}\right)^{\rm cont}
+\left[\frac{\epsilon_Q^{\rm line}}{\epsilon_I^{\rm line}}
\frac{S^{\rm line}}{B} - 
\frac{\eta_Q^{\rm line}}{\eta_I^{\rm line}}\right]
\frac{\eta_I^{\rm line}}{\eta_I^{\rm cont}},    \label{14}
\end{equation}
where $S^{\rm line}$ and $B$ are the line and continuum source functions,
respectively.
The expression between brackets in Eq.~(\ref{14}) is easy to evaluate 
once the $\rho^0_0$ and $\rho^2_0$ elements of the levels involved are known {because}
$\epsilon_i^{\rm line}$ and $\eta_i^{\rm line}$ are given 
in Eqs.~(\ref{9})-(\ref{11b}) above {and, assuming
that the continuum source function is the Planck function, we have}
\begin{equation}
\frac{S^{\rm line}}{B}=\left(\frac{2J_\ell+1}{2J_u+1}\right)^{1/2}
\frac{\rho^0_0(u)}{\rho^0_0(\ell)}({\rm e}^{h\nu/k_{\rm B}T}-1),
\end{equation}
where we assume a temperature $T=5800$~K.
The ratio ${\eta_I^{\rm line}}/{\eta_I^{\rm cont}}$ in Eq.~(\ref{14})
depends critically on the details of the line formation 
and cannot be calculated without a true radiative transfer treatment.
Consistently with our simplified transfer model, we assume 
that this ratio is given by some
arbitrary value $\alpha$, identical for all the lines, and we
take the value that better match the observations.
More generally, $\alpha$ also accounts for any systematic effect 
(identical for all the lines), not considered
in our modeling like, for instance, the lower polarization degree  
for on-disk observations with respect to the tangential limit, 
and the broadening of line profiles.

Figure 1 shows the line core polarization in 26 lines of the 
Ce {\sc ii} solar spectrum (shadow bars) as reported by
Gandorfer (2000, 2002; to be referred to as the Atlas hereafter).
The lines have been selected as those showing a clear polarization
peak, either above the continuum or as a depolarization feature
in the Atlas. We have avoided lines with severe blends
that do not allow {a clear assignment of the} polarization signal. In particular,
we have avoided lines having blends 
with species that are known to have notable scattering 
polarization signals (e.g., Ti~{\sc i}, Cr~{\sc i}, molecules), 
or lines that lie on a highly structured polarization background 
(e.g., on the wings of the H- and K-lines of Ca {\sc ii}).
We point out, however, that sometimes it is 
possible to ascribe clearly some $Q/I$ signal to a line despite 
an intensity blend. Those lines have
also been included in the study.

The heavy line in Fig.~1 shows the polarization signal in each transition
calculated according to the analysis above (Eqs.~(\ref{see1})-(\ref{14})).
{Our simplified scattering polarization modeling} is able to reproduce the sign of most of the 
selected lines and roughly the relative $Q/I$ signals among lines.
A quantitative evaluation of the agreement between the 
observed polarization pattern and the theoretical estimate 
may be given through the correlation factor 
$r=\sum p_iq_i/(\sum p_i^2\sum q_i^2)^{1/2}$, where $p_i$ and $q_i$
correspond to the observed and theoretical 
fractional polarization in the $i$-th line, referred to the continuum.
For the case shown in Fig.~1, $r=0.72$.

A better modeling of the observed pattern requires to
include depolarizing collisions and turbulent magnetic fields.
A detailed treatment of depolarizing collisions is not possible
because of the lack of data. Some progress is indeed going on
on the theoretical side (e.g., Derouich et al. 2003, 2004; Kerkeni 2002), 
but no reliable calculations exist (to the authors' knowledge) of 
depolarizing rates in rare earth atoms.
Nonetheless, much qualitative understanding on their role may be gained
by proceeding as in Papers {\sc i}; we just consider one and the 
same depolarizing collisional rate $D$ for all the statistical 
tensors with $K\neq 0$ in all the levels 
and we add a rate of the form 
$-D \rho^K_0(i)$
to the statistical equilibrium equation of each $\rho^K_0(i)$ element
with $K\neq 0$.

\begin{figure}
\centerline{\epsfig{figure=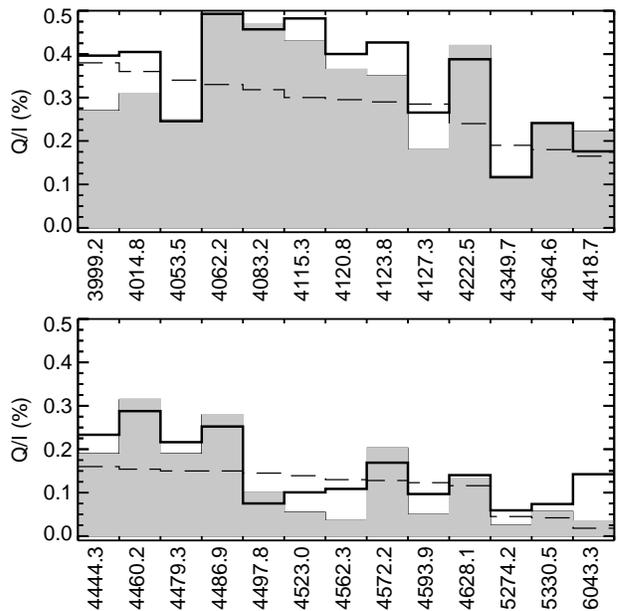, width=9cm}}
\caption{Same as Fig.~1 with a common depolarizing rate $D=10^6$~s$^{-1}$ 
for all the levels of the Ce~{\sc ii} model, and $\alpha=0.12$.}
\end{figure}

Figure~2 shows the fractional
polarization calculated assuming a 
depolarizing rate $D=10^6$~s$^{-1}$ for all the atomic levels (heavy line).
The agreement with the observed polarization pattern 
(shadow bars) is better than in the $D=0$ case (cf. Fig.~1); 
quantitatively, this is reflected in a larger correlation factor $r=0.83$.
However, the agreement does not improve further 
for increasingly larger depolarizing rates but, on the contrary, it 
degrades, which
puts an upper limit to the depolarizing collisional rate at about
$\sim 10^6$~s$^{-1}$.

A depolarizing rate $D=10^6$~s$^{-1}$ leaves 
the atomic polarization of the lowermost levels (ground and metastable) of Ce {\sc ii}
at about one fifth of the atomic polarization in the absence of 
depolarizing processes.
This is interesting because
{a microturbulent} magnetic field may also depolarize an atomic level
whenever its corresponding Larmor frequency $\nu_L$ is of the order of the 
relaxation rates of the level (i.e., the inverse of its mean life-time $\tau$)
or larger.
But, differently from the case of collisions, the depolarization
{\em saturates} when $\nu_L\gg 1/\tau$, and the atomic polarization 
of the level remains at a constant value
that is one fifth that of the non magnetic case (see Appendix).
Therefore, the results in Fig.~2 are consistent with the 
presence of a turbulent magnetic field
in the saturation regime of the Hanle effect for the lower levels.
The lowermost levels of our Ce~{\sc ii} model atom have mean life times
$\tau\approx$1-8$\times 10^{-6}$~s; the turbulent magnetic field 
required for reaching the saturation regime of the lowermost levels of 
Ce~{\sc ii} is hence of the order of 0.1-0.8~gauss.

On the other hand, the magnetic field cannot be so strong to saturate
also the upper levels of the transitions, or the fractional polarization
pattern would be exactly as in the non magnetic case but reduced to 
one fifth, and therefore, $r$ would degrade again to 0.72.
The upper levels of the transitions considered in this study have
mean lifetimes $\tau\approx 4$-6$\times 10^{-9}$~s; most of these levels
saturate with fields $B\gtrsim$~200~gauss, all of them when 
$B\gtrsim$~300~gauss.

\section{Conclusions}

Rare earths are atoms with a complex level structure 
whose scattering polarization patterns
cannot be understood using simple two level approximations.
Moreover, in astrophysical environments (and in the solar atmosphere in 
particular), many transitions are
excited simultaneously and the population and atomic polarization of a level depend  
on the balance between all the absorption and emission processes
involving the level.

Here, we have considered the generation and distribution of atomic
alignment in a multilevel atomic model of Ce {\sc ii} 
when it is illuminated by a solar phostospheric-like
radiation field. We have seen that the ensuing polarization of the 
scattered light is in good agreement with
the observed polarization
pattern of Ce {\sc ii}.
In particular, our model is capable of explaining the strongest polarized signals, 
as well as the sign of the polarization signal observed in 
most of the lines selected in this study.

With a simple analysis of the role of depolarizing collisions 
we have shown that the best agreement with observations is achieved 
at depolarizing rates $D=10^6$~s$^{-1}$, thus establishing
an upper limit to these collisions. 
In this regime, the lowermost levels of ionized cerium show
an atomic polarization that is of the order of one fifth of that which would
be present in the absence of 
any depolarizing mechanisms.
We interpret this fact as an indication of the presence of 
{volume-filling microturbulent magnetic field of strength}
$B\gtrsim 0.8$~gauss (the 
saturation regime of the Hanle effect for the lowermost levels).
Since in the presence of a magnetic field strong enough to saturate also the 
upper levels the relative polarization of the lines would match observations
exactly as in the non magnetic case (which is not the best fit), 
we conclude that the magnetic field
is not saturating the upper levels and hence the magnetic field
must be $B\lesssim$~200--300~gauss.
  
In summary, we have given a theoretical interpretation of the
second solar spectrum of Ce {\sc ii} and shown that it is a
promising diagnostic tool to probe the magnetism and thermodynamics
of the solar atmosphere.
However, we cannot proceed much further with the elementary methods 
used in this paper.

Two are the main limitations of this research that must be addressed 
next. 
First, the full radiative transfer problem must be solved, thus
allowing for different heights of formation of the lines.
Second, the role of the magnetic field must be included consistently 
beyond the analogy with depolarizing collisions used here.
The results of these investigations will be presented in forthcoming
papers.

\begin{acknowledgements}

{We thank Roberto Casini for the careful reading of the manuscript and J. O. Stenflo for his comments to improve it.
One of the authors (RMS) gratefully acknowledges partial support
from the University of Florence via an Assegno di Ricerca. 
This work has also benefited from finantial support through project AYA2004-05792 of the Spanish Ministerio de Educaci\'on y Ciencia and 
the European Solar Magnetism Network HPRN-CT-2002-00313.
This research has made use of NASA's Astrophysics Data System.}

\end{acknowledgements}

\appendix

\section{Saturation Regime of the Hanle Effect from a
Microturbulent Field}

When a magnetic field is present in an otherwise axisymmetric enviroment, 
it is not possible, in general,
to characterize the excitation state of the atoms only 
with the populations of the individual magnetic sublevels (or 
equivalently, 
through 
the statistical tensors $\rho^K_0$).
Taking the quantization axis along the direction
of the magnetic field, the radiation field is characterized through the $J^K_Q$
components ($K=0, 2$, $Q=-K, ..., K$), 
where the $Q\ne 0$ elements are azimuthal
averages of the intensity of the radiation field  
(e.g., Landi Degl'Innocenti \& Landolfi 2004).
Correspondingly, the atomic excitation is described by the 
larger set of statistical tensors $\rho^K_Q$,
($K=0, 2, 4, ..., K_{\rm max}$, $Q=-K, ..., K$; $K_{\rm max}=2J$ or $2J-1$
for integer or half-integer $J$, respectively), 
where the $Q\ne 0$ elements correspond to
quantum coherences between magnetic sublevels
(e.g., Fano 1957; Landi Degl'Innocenti \& Landolfi 2004).
If the radiation field satisfies the condition $J^K_Q/J^0_0\ll 1$
(for, $K, Q\neq 0$), 
which is typically the case in stellar atmospheres, 
then $\rho^K_Q/\rho^0_0\ll 1$ (for, $K, Q\neq 0$), and a perturbative
treatment of the statistical equilibrium equations can be performed 
(weak anisotropy approximation; see Landi Degl'Innocenti 1984;
Landi Degl'Innocenti \& Landolfi 2004, Sect. 10.13).
Then, to first order, the statistical equilibrium equations
in the magnetic field reference system do not 
couple the $\rho^K_Q$ components with $\rho^{K'}_{Q'}$ components 
having $Q\neq Q'$.

Let ${\cal V}$ and ${\cal M}$ be the reference systems with the 
quantization axis directed along the vertical and the magnetic field, 
respectively;
let R be the rotation taking ${\cal V}$ into ${\cal M}$, and ${\cal D}^K_{QQ'}(R)$
($Q, Q'=-K, ..., K$) the corresponding rotation matrix 
(e.g., Brink \& Satchler 1968).
Let $[\rho^2_0]^\circ$ be the alignment of a level in the reference system
${\cal V}$, for zero magnetic field. 
In the reference system ${\cal M}$, 
it turns out that 
$[\rho^2_0]^{\rm m}={\cal D}^2_{00}(R)[\rho^2_0]^\circ$.
This is true even if the  magnetic field is not zero, since it does
not alter 
$[\rho^2_0]^{\rm m}$ directly, and $[\rho^2_0]^{\rm m}$  is decoupled from 
other $[\rho^2_Q]^{\rm m}$ components with $Q\neq 0$ (see above).
Moreover, if the Larmor frequency $\nu_L$ is much larger than the radiative 
rates ${\cal R}_i$ of the level, this is the only non zero density matrix element because 
$[\rho^K_Q]^{\rm m}\rightarrow 0$ for components $Q\ne 0$.
Finally, passing from ${\cal M}$ to ${\cal V}$,
$[\rho^K_Q]={\cal D}^K_{Q0}(R)^*{\cal D}^K_{00}(R)[\rho^2_0]^\circ$.
Averaging over all the directions of the turbulent field, 
$[\rho^2_Q]=\delta_{Q0}[\rho^2_0]^\circ/5$.
Therefore, the alignment of a level in the saturated Hanle effect 
regime ($\nu_L\gg {\cal R}_i$), is 1/5 of its 
'non magnetic' value. 

Now, suppose there is magnetic field regime such that
the atomic levels can be classified in two types; type A: levels  
in the saturated regime and type B: levels in the regime $\nu_L\ll {\cal R}_i$.
The alignment of levels in set A is 1/5 of the value obtained by solving
Eqs.~(\ref{see1})-(\ref{see4}).
The alignment of the levels in set B is obtained solving  
a restricted system of equations under this constraint.
Moreover, it can also be shown that 
the alignment of the levels in set B is the weighted average
(with weights 1/5 and 4/5, respectively), of the value obtained by solving
Eqs.(\ref{see1})-(\ref{see4}), and of the value obtained by solving
the same equations under the constraint of complete depolarization of the 
levels of set A.

The rationale for this problem is the following. 
Because of the large life-time difference between metastable and excited
levels, the lower- and upper-level Hanle effect regimes are usually
well separated 
($\tau_\ell/\tau_u\sim A_{u\ell}/B_{\ell u}J^0_0 \sim 100$ for two levels
conected by an allowed visible transition). 
Therefore, there is a regime of magnetic fields in which metastable levels
are completely saturated while excited levels remain essentially unaffected
by the magnetic field. 
In some astrophysical enviroments the microscopic 
fields created by the motion of nearby ions may be large enough to 
saturate the Hanle effect of metastable levels. 
From a physical point of view, the zero (macroscopic) field case then 
corresponds to the solution of the problem just stated.


\begin{thebibliography}{}
\bibitem[2005]{}
Asplund, M., Grevesse, N., \& Sauval, A. J. 2005, 
in ASP Conf. Ser. 336, 
Cosmic Abundances as Records of Stellar Evolution and Nucleosynthesis,
eds. Thomas G. Barnes {\sc iii} and Frank N. Bash(San Francisco: ASP), 
p.~25; arXiv:astro-ph/0410214v2
\bibitem[]{}
Brink, D. M., \& Satchler, G. R. 1968, Angular Momentum (Oxford: Clarendon
        Press)
\bibitem[2000]{cox}
Cox, A.N. 2000, Allen's Astrophysical Quantities 4th ed. (New York: 
	Springer Verlag and AIP Press)
\bibitem[]{}
Derouich, M., Sahal-Br\'echot, S., Barklem, P. S., \& O'Mara, B. J. 2003,
        A\&A, 404, 763
\bibitem[]{}
Derouich, M., Sahal-Br\'echot, S., \& Barklem, P. S. 2004,
        A\&A, 426, 707
\bibitem[1980]{}
van Diest, H. 1980, A\&A, 83, 378
\bibitem[]{}
Fano, U. 1957, Rev. Mod. Phys., 29, 74
\bibitem[2000]{gandorfer}
Gandorfer, A. 2000, The Second Solar Spectrum: A high spectral resolution 
polarimetric survey of scattering polarization at the solar limb in graphical 
representation. Volume {\sc i}: 4625~\AA\  to 6995~\AA\ (Z\"urich: vdf 
Hochschulverlag AG an der ETH)
\bibitem[2000]{gandorfer}
Gandorfer, A. 2002, The Second Solar Spectrum: A high spectral resolution 
polarimetric survey of scattering polarization at the solar limb in graphical 
representation. Volume {\sc ii}: 3910~\AA\  to 4630~\AA\  (Z\"urich: vdf 
Hochschulverlag AG an der ETH)
\bibitem[1981]{}
Irwin, A. W. 1981, ApJS, 45, 621
\bibitem[]{}
Kerkeni, B. 2002, A\&A, 390, 783
\bibitem[1984]{gidi84}
Landi Degl'Innocenti, E. 1984, Sol. Phys., 91, 1
\bibitem[]{}
Landi Degl'Innocenti, E. \& Landolfi, M. 2004, Polarization in Spectral Lines
        (Dordrecht: Kluwer)
\bibitem[]{}
Manso Sainz, R., \& Landi Degl'Innocenti, E. 2002, A\&A, 394, 1093 (Paper {\sc i}a)
\bibitem[]{}
Manso Sainz, R., \& Landi Degl'Innocenti, E. 2003, in ASP Conf. Ser. 307,
        Solar Polarization 3, eds. J. Trujillo Bueno and
	J. S\'anchez Almeida (San Francisco: ASP), 425  (Paper {\sc i}b)
\bibitem[1978]{martin}
Martin, W. C., Zalubas, R., \& Hagan, L. 1978, 
        Atomic Energy Levels---The Rare-Earth Elements,
	Natl. Stand. Ref. Data Ser., Natl. Bur. Stand. (U.S.) monograph 60.
\bibitem[2000]{}
Palmeri, P., Quinet, P., Wyart, J.-F., \& Bi\'emont, E. 2000,
        Phys. Scripta, 61, 323
\bibitem[1996]{stenflo1}
Stenflo, J.O., \& Keller, C.U. 1996, Nature, 382, 588
\bibitem[1997]{stenflo2}
Stenflo, J.O., \& Keller, C.U. 1997, A\&A, 321, 927
\bibitem[2003]{trujillo}
Trujillo Bueno, J. 2003, in Solar Polarization 3, ed. J. Trujillo Bueno \& J. S\'anchez Almeida, ASP Conf. Ser. Vol. 307, 407
\bibitem[]{}
Vernazza, J. E., Avrett, E. H., \& Loeser, R. 1976, ApJS, 30, 1
\bibitem[]{}
Zhang, Z. G., Svanberg, S., Zhankui Jiang, Palmeri, P., Quinet, P., 
        \& Bi\'emont, E. 2001, Phys. Scripta, 63, 122
\end{thebibliography}
\end{document}